\documentstyle[amssymb,twocolumn,prb,aps]{revtex}  
\input{epsf.sty}

\begin{document}   
\draft 
\title{Temperature-dependent electronic structure and ferromagnetism in the
  $d=\infty$ Hubbard model \\ studied by a modified perturbation theory}
\author{T.~Wegner~\thanks{e-mail: wegner@physik.hu-berlin.de}, M.~Potthoff
  and
  W.~Nolting} \address{Institut f\"ur Physik,\\
  Humboldt-Universit\"at zu Berlin, Invalidenstra{\ss}e 110, D-10115 Berlin,
  Germany}
%
\maketitle
\begin{abstract}                
  The infinite-dimensional Hubbard model is studied by
  means of a modified perturbation theory.  The approach reduces to the
  iterative perturbation theory for weak coupling. It is exact in
  the atomic limit and correctly reproduces the dispersions and the weights of
  the Hubbard bands in the strong-coupling regime for arbitrary fillings.
  Results are presented for the hyper-cubic and an fcc-type
  lattice. For the latter we find ferromagnetic
  solutions. The filling-dependent Curie temperature is compared with the
  results of a recent Quantum Monte Carlo study.
\end{abstract}


Correlations among itinerant electrons are responsible for various interesting
phenomena like spontaneous magnetic order, the metal-insulator (Mott)
transition and high-temperature super-conductivity. One of the simplest but
non-trivial models describing correlated electrons on a lattice is the Hubbard
model~\cite{Hub63}. Studying the Hubbard model in the limit of
infinite spatial dimensions~\cite{MV89,Vol93} $d$ is of special importance for
the construction of a dynamical mean-field theory~\cite{Vol93}. For $d=\infty$
the Hubbard model is considerably simplified but nevertheless remains
non-trivial: It becomes equivalent to an effective impurity
problem and can be mapped~\cite{GK92,Jar92} onto the
single-impurity Anderson model (SIAM), for example. For the
latter numerically exact solutions can be obtained by Quantum Monte Carlo
(QMC) calculations. On the other hand, it may be helpful to have
an analytical (but approximative) expression for the self-energy at one's
disposal, which recovers the exact QMC results as reliable as possible. This
allows a direct calculation of dynamic quantities on the real energy axis for
finite temperatures as well as for $T=0$. Furthermore, valuable hints for the
approximative solution of more complicated lattice models not accessible to
QMC may be obtained.

For the symmetric case
the iterative perturbation theory~\cite{GK92,GK93} (IPT)
is known to give a rather realistic description of the Mott transition. The
IPT employs the self-consistent mapping onto the SIAM which is solved by means
of second-order perturbation theory around the Hartree-Fock solution
(SOPT-HF). To extend the IPT to non-symmetric cases, Kajueter and
Kotliar~\cite{KK96} proposed an interpolating expression for the self-energy
which for arbitrary band-fillings reproduces the atomic as well as the
weak-coupling limit.

For the study of ferromagnetism special attention has to be paid to the
strong-coupling regime. Fortunately, the $1/U$ perturbation theory of Harris
and Lange~\cite{HL67} provides rigorous results for $U\rightarrow \infty$: In
the first non-trivial order beyond the atomic limit, the average dispersions as
well as the weights of the two dominating Hubbard bands are known exactly
\cite{HL67}. One possibility to account for these
strong-coupling results in the construction of an analytical expression for
the self-energy, is to ensure that the moments
\begin{equation}
\label{moments}
  M^{(n)}_{{\bf k}\sigma} = \frac{1}{\hslash} \int \! dE \, E^n A_{{\bf k}
      \sigma}(E)
\end{equation}
of the resulting spectral density $A_{{\bf k}\sigma}(E)$ are correct up to
$n=3$. These moments can be calculated exactly by $M^{(n)}_{{\bf k}\sigma}
 = \langle [{\cal L}^n c_{{\bf k}\sigma} , c_{{\bf k}\sigma}^\dag ]_+
 \rangle $,
where ${\cal L} = [\,\cdot\,,{\cal H}]_-$. At each
${\bf k}$ point of the Brillouin zone the first four moments $n=0,\ldots,3$
provide four pieces of information that in the strong-coupling regime determine
the dispersions and the weights of the two Hubbard bands. This moment approach
has been successfully employed beforehand to improve upon the Hubbard-I
solution~\cite{NB89}.

The approach of Kajueter and Kotliar~\cite{KK96} reproduces the correct
moments up to $n=2$. The $n=3$ moment, however, may be of particular
importance in the context of ferromagnetism since it involves a higher-order
correlation function (``band shift''),
\begin{equation}
  \label{bandshift}
  B_{\sigma} = T_{ii} +
  \sum_{j\ne i} T_{ij} \frac{\left\langle c_{i\sigma}^\dagger c_{j\sigma} 
  (2 n_{i-\sigma} - 1)\right\rangle}{\langle n_{i\sigma}\rangle\left(1-\langle
    n_{i\sigma} \rangle\right)}\;,
\end{equation}
the (possible) spin-dependence of which is known to favour magnetic order and
thereby decisively influences the magnetic phase diagram \cite{NB89}.

Based on these considerations the present authors have proposed an improvement
of the Kajueter-Kotliar approach recently \cite{PWN97} which is correct up to
$n=3$ (hereafter referred to as ``modified perturbation theory'', MPT).  The
main purpose of this paper is to study the effects of the band shift
$B_\sigma$ in the paramagnetic as well as in the ferromagnetic phase. We
compare our results with corresponding findings of previous QMC studies.

Let us briefly recall the essentials of the theory. Details can be
found in Ref. \onlinecite{PWN97}.
In the $d=\infty$ Hubbard model the self-energy is a local
quantity~\cite{MV89,Vol93}. Via the mapping onto the effective
impurity problem, it can be regarded as the self-energy of the SIAM provided
that the hybridisation function of the SIAM is suitably chosen.  According to
Kajueter and Kotliar~\cite{KK96}, we consider the following ansatz for the
self-energy of the SIAM:
\begin{equation}
  \label{ansatz}
  \Sigma_{\sigma}(E)=U n_{-\sigma}+
  \frac{a_\sigma\Sigma_{\sigma}^{\rm (SOC)}(E)}
     {1-b_\sigma\Sigma_{\sigma}^{\rm (SOC)}(E)}\,,
\end{equation}
where $\Sigma_{\sigma}^{\rm (SOC)}(E)$ is the second-order ($U^2$) contribution
(SOC) to the self-energy within the SOPT-HF, while $a_\sigma$ and $b_\sigma$ are
free parameters which will be chosen such that the moments~(\ref{moments}) are
correct up to $n=3$. This requires $a_\sigma$
and $b_\sigma$ to be~\cite{PWN97}:
\begin{equation}
  \label{asi}
  a_\sigma = \frac{n_{-\sigma} \left(1-n_{-\sigma}\right)}
  {n_{-\sigma}^{\rm (HF)}\left(1-n_{-\sigma}^{\rm (HF)}\right)}\\
\end{equation}
\begin{equation}
  \label{bsi}
  b_\sigma = \frac{B_{-\sigma}-B_{-\sigma}^{\rm (HF)}-
     (\mu-\widetilde{\mu}_\sigma)
    +U\left(1-2n_{-\sigma} \right)} {U^2
    n_{-\sigma}^{\rm (HF)}\left(1-n_{-\sigma}^{\rm (HF)}\right)}.
\end{equation}
$n_{\sigma}$ denotes the occupancy of the
impurity level, $n_{\sigma}^{\rm (HF)}$ is its Hartree-Fock value,
and $\widetilde{\mu}_\sigma$ is a fictitious chemical potential that appears in
the definition of the Hartree-Fock Green's function. The parameter
$\widetilde{\mu}_\sigma$ is fixed by imposing the condition
$n_{\sigma}^{\rm (HF)}=n_{\sigma}$ (for further discussions on this point see
Ref.\onlinecite{PWN97}).  Finally, $B_{\sigma}^{\rm (HF)}$ is the Hartree-Fock
value of the band shift $B_\sigma$. All expectation values in the theory
can be expressed in terms of the spectral density and thus can be determined
self-consistently~\cite{PWN97}.

The modified perturbation theory (MPT) reduces to the IPT for small $U$ and is
exact in the atomic limit.  By construction it is fully consistent with the
rigorous results of Harris and Lange for $U\rightarrow \infty$ and yields the
correct
moments of the spectral density up to $n=3$ for arbitrary $U$. The approach of
Kajueter and Kotliar~\cite{KK96} is recovered if $B_\sigma$ and $B_\sigma^{\rm
  (HF)}$ are set to their atomic limit values (i.~e. $B_\sigma = B_\sigma^{\rm
  (HF)} = T_{ii}\equiv 0$).  Note, however, that in Ref.~\onlinecite{KK96} the
fictitious chemical potential $\widetilde{\mu}_\sigma$ has been used to
enforce the Luttinger theorem. This implies that
the theory is intrinsically limited to $T=0$ (see Ref.~\onlinecite{PWN97}).
The more unproblematic condition $n_\sigma^{\rm(HF)} = n_\sigma$ which is
chosen here to fix $\widetilde{\mu}_\sigma$ has to be preferred since it
allows to perform finite-temperature calculations, too. The results
for $T=0$ and near half-filling show that this choice is still consistent with
the Luttinger theorem (see below).

Let us first consider the paramagnetic Hubbard model on the hyper-cubic lattice
away from half-filling ($n<1$). The bloch density of states (BDOS) is given
by~\cite{MV89} $\rho_0(E)=\exp(-E^2)/\sqrt{\pi}$.
Fig.~1 shows the density of states (DOS) for $U=4$,
$k_BT=0.138$ and different band-fillings obtained within the MPT.  The
comparison with the QMC results of Jarrell and Pruschke~\cite{JP93} shows that
the MPT qualitatively yields the correct results as concerns the shifts as
well as the changes in height and width of the high-energy charge-excitation
peaks in the spectrum with varying filling. Even for temperatures $T\neq 0$
the Kondo-type
resonance shows up at $E\approx \mu$. This low-energy feature is more
pronounced in the QMC spectra compared with the MPT.
We then investigated the effect of taking into account the $n=3$ moment: For
any filling considered, it turns out that there are only minor changes in the
spectra (which would hardly be visible on the scale in
fig.~1) when setting $B_\sigma=B_\sigma^{\rm (HF)}$. For the
paramagnet we conclude that the effects introduced by the band shift are
rather unimportant.

\begin{figure}
    \begin{center}
    \mbox{} \epsfxsize=0.8\linewidth \epsffile{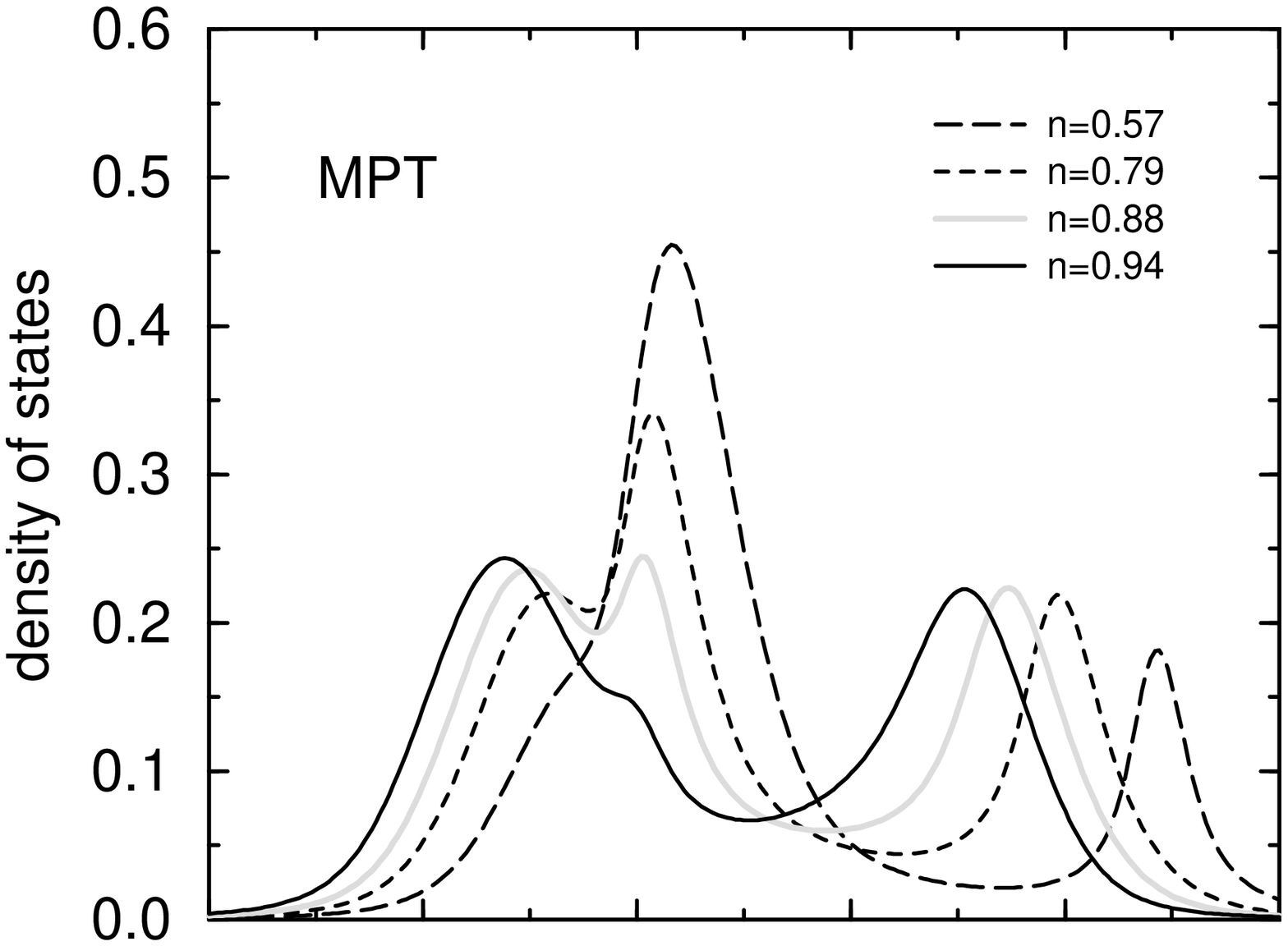}
    \vspace{-1cm} \mbox{} \epsfxsize=0.8\linewidth
    \epsffile{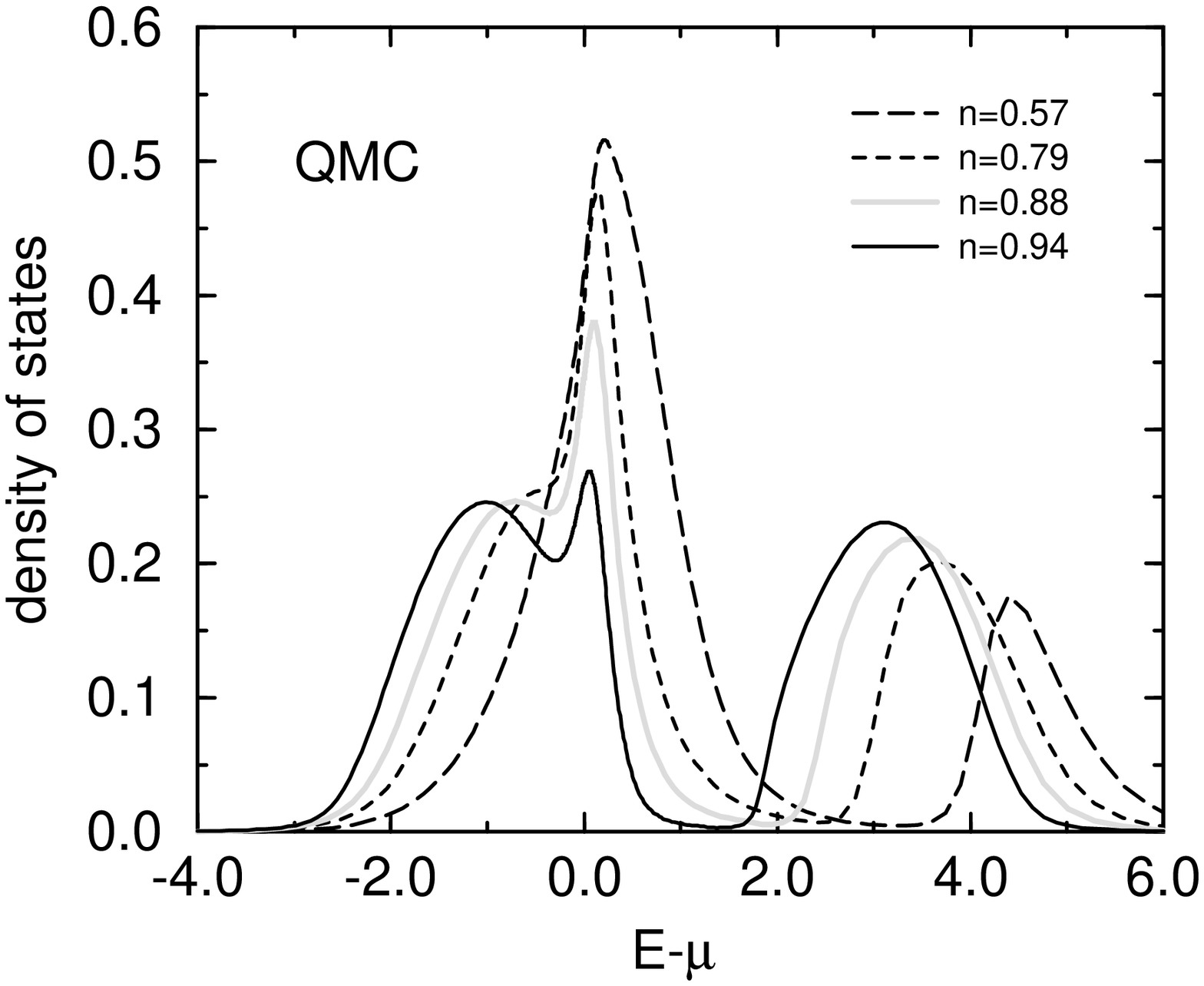}
    \caption{
   	Densities of states for $U=4$, $k_BT=0.138$ and different 
      fillings on the hc lattice. Results of the modified perturbation theory
      (MPT) in comparison with Quantum Monte Carlo (QMC) results from Jarrell
      and Pruschke~\protect\cite{JP93}.}
    \end{center}
    \label{fig:mptqmc}
\end{figure}
\begin{figure}
    \begin{center}
    \mbox{} \epsfxsize=0.8\linewidth \epsffile{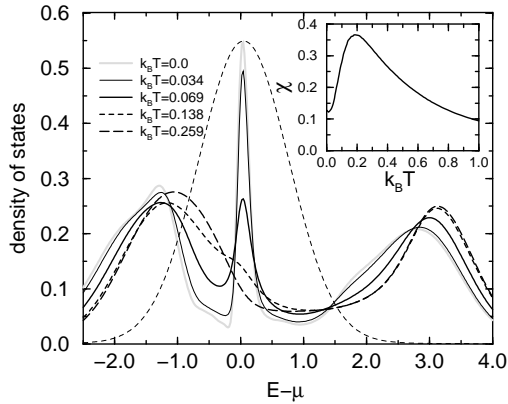}
    \caption{Densities of states for $U=4$, $n=0.94$ and different
      temperatures. Thin dashed line: $U=0$.  Inset: uniform static
      susceptibility $\chi$ vs temperature.}
    \end{center}
    \label{fig:kondo}
\end{figure}

Fig.~2 shows the MPT-spectra for $U=4, n=0.94$ and different
temperatures. We have also calculated the uniform static susceptibility in the
paramagnetic phase by applying an infinitesimally small field:
$\chi=\left.\frac{\partial}{\partial H}
(n_\uparrow-n_\downarrow)\right|_{H\rightarrow 0}$.  As can
be seen in fig.~\ref{fig:kondo} the Kondo peak starts growing when the
susceptibility differs from the Curie-law. The local effective magnetic
moment ($\mu_{\rm{eff}}^2\propto T\chi(T)$) is quenched as 
$T\rightarrow 0$. This effect has been termed ``collective single-band
Kondo-effect''~\cite{PJF95}.
Furthermore one can see in fig.~\ref{fig:kondo} by comparing the
$T=0$-spectrum with the free ($U=0$) spectrum that the DOS is unrenormalised
at the Fermi edge, $\rho_\sigma(\mu)=\rho_0(\mu_0)$, which is an equivalent
formulation of the Luttinger theorem in infinite dimensions~\cite{MHa89}. Let
us stress once more that contrary to Ref.~\onlinecite{KK96} this has not been
enforced via the choice for the parameter $\widetilde{\mu}_\sigma$.
For fillings well below half-filling $\rho_\sigma(\mu)$ is found 
to be slightly lower than $\rho_0(\mu_0)$.
We did not find ferromagnetic solutions on the hc lattice. The susceptibility
never diverged for $U\leq6$. This is in agreement with the results in
Ref.~\onlinecite{JP93}. For $U=\infty$, however, a region of
non-vanishing ferromagnetic polarisation is obtained within the non-crossing
approximation~\cite{OPK97}.

On an fcc-type lattice ferromagnetism has been found recently within QMC by
Ulmke~\cite{Ulm97}. The high weight in the BDOS~\cite{Ulm97}
$\rho_0(E)=\exp(-(1+\sqrt{2}E)/2)/\sqrt{\pi (1+\sqrt{2}E)}$
at the lower band edge favours ferromagnetic
order~\cite{VBHKSU97}. Stable ferromagnetic solutions are likewise predicted
by the MPT in a wide region of the phase diagram.  Fig.~3
shows the magnetisation $m=n_\uparrow - n_\downarrow$ for $U=4$ as a function of
temperature for different band-fillings. For $n<0.67$ we observe second-order
phase transitions. The magnetisation curves are continuous and terminate at
the respective Curie temperature $T_C$.
The inverse static susceptibility obeys a Curie-Weiss law for high
temperatures. For lower temperatures with $T > T_C$ a slight curvature  
($\partial^2 \chi^{-1} / \partial T^2 > 0$) is observed
(not visible in fig.~3). $\chi^{-1}$ vanishes at 
$T=T_C$. The critical exponent of the susceptibility turns out to be
$\gamma\approx 1$ as expected. We encountered numerical difficulties to
obtain an accurate value for $\gamma$.

For $n \ge 0.67$ the phase transitions are of first order. A finite
magnetisation is still found for temperatures above the zero $T_1$
of $\chi^{-1}$ (see fig.~3 for $n=0.67$).
At a temperature $T=T_2>T_1$ the magnetisation non-continuously
drops to zero. Between $T_1$ and $T_2$ there is a second 
ferromagnetic solution with a magnetisation (not shown) that 
vanishes at $T_1$ and that coalesces with the plotted magnetisation
curve of the first solution at $T_2$. Decreasing $n$ also results
in a stronger curvature of $\chi^{-1}$ near $T_1$ while at high 
temperatures we still have Curie-Weiss behaviour. Finally, for 
$n \ge 0.69$, $\chi^{-1}$ remains positive for all temperatures. 
This scenario is well known from previous studies of the Hubbard 
model in conserving~\cite{HC94} as well as non-conserving 
approximations~\cite{NBB91}. For $n \ge 0.67$ the true Curie 
temperature $T_C$ (with $T_1 < T_C < T_2$) can be found e.~g.~by 
means of the Maxwell construction considering the $H$-$m$ isotherms 
at different temperatures. On the other hand, we cannot exclude that 
the first-order transition is an artefact of our approximation. 

\begin{figure}
    \begin{center}
    \mbox{} \epsfxsize=0.8\linewidth \epsffile{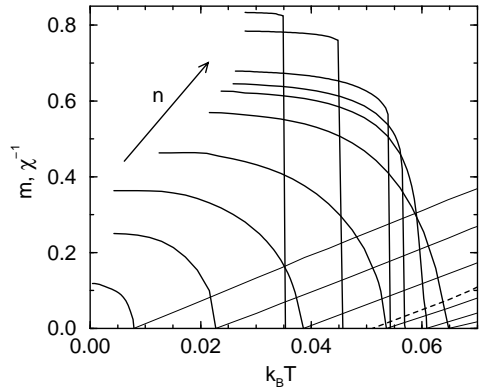}
    \caption{Thick lines: magnetisation for the fcc-lattice, $U=4$ and
      $n=0.2,0.3,0.4,0.5,0.6,0.65,0.67,0.7,0.8,0.85$ vs temperature. Thin
      lines: corresponding inverse static susceptibility $\chi^{-1}$ for
      $n=0.2,\ldots,0.65$. Dashed line: $n=0.67$.}
    \end{center}
    \label{fig:fcc_m_n_T}
\end{figure}
The filling-dependence of $T_C$ (for $n \ge 0.67$: $T_1$) can be 
compared with the available numerically exact QMC results from 
Ref.~\onlinecite{Ulm97} where the Curie temperature has been obtained by linear 
extrapolation of the inverse susceptibility. As can be seen in 
fig.~4, the MPT result reasonably agrees with the QMC data. In 
particular, there is remarkable agreement with respect to the 
maximum $T_C$ and the corresponding filling.
Contrary to the paramagnetic phase, there is a strong effect of the
(spin-dependent) band shift $B_\sigma$ for the ferromagnet.  Both, the range
of existing ferromagnetic solutions as well as the values for $T_1$ and $T_C$,
are strongly affected if we set $B_\sigma = B_\sigma^{\rm (HF)}$. We find that
taking into account correctly the $n=3$ moment tends to improve the agreement
with the QMC results (see fig.~4).

There are no hints for non-Fermi-liquid behaviour in all our 
calculations. The imaginary part of the self-energy vanishes 
quadratically at $E=\mu$ in the para- as well as in the 
ferromagnetic solutions for $T=0$, even at the quantum-critical
point $n_c \approx 0.16$ (for $U=4$).

The evolution of the spin-dependent spectral density with increasing
temperature is shown in fig.~5 for $n=0.6$. Since the self-energy
is ${\bf k}$-independent, the spectral density depends on ${\bf k}$ via the
Bloch dispersion $\epsilon({\bf k})$ only. For the calculation of the density
of states which is also shown in fig.~5, the spectral density has
to be weighted with the BDOS:
\begin{equation}
  \label{qdos}
  \rho_{\sigma}(E)=\frac{1}{\hslash} \int \! d\epsilon({\bold k}) \,
  \rho_0(\epsilon({\bold k})) A_{\epsilon({\bold k})\sigma}(E-\mu)\;.
\end{equation}
At the temperature $k_BT=0.022$, where the system is nearly fully polarised
($m=0.57$), the upper Hubbard band in the $\uparrow$-spectrum is missing
because
there are not enough interaction partners. The same spectrum without the
narrow Kondo-type resonance (at $E\approx 0$) is expected in the
non-interacting case. The $\downarrow$-spectrum shows the lower as well as the
upper
Hubbard band. When increasing the temperature (second row, $k_BT=0.058$,
$m=0.32$) the upper Hubbard band in the $\uparrow$-spectrum comes into
existence.
The resonance smears out and starts growing in the $\downarrow$-spectrum.
At $k_BT=k_BT_C=0.065$ (third row) where $\uparrow$-~and $\downarrow$-spectra
coincide, it
is still visible. Finally, we notice that there is a strong transfer of
spectral weight between the Hubbard bands with increasing temperature. In the
$\uparrow$-channel the upper one gains weight at the cost of the lower, in the
$\downarrow$-channel the situation is reverse.
\begin{figure}
   \begin{center}
    \mbox{} \epsfxsize=0.8\linewidth \epsffile{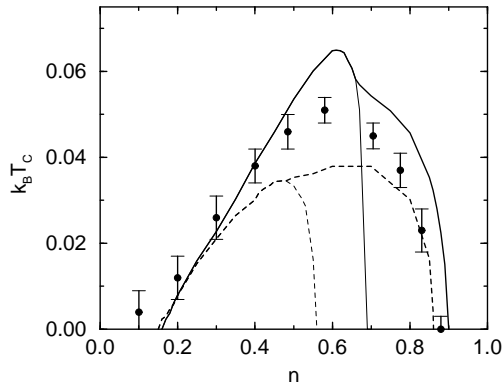}
    \caption{Filling dependence of the Curie-temperature $T_C$ ($T_2$ for $n
    	\ge 0.67$) for $U=4$ on
      the fcc-lattice. Solid lines: MPT. Dashed lines: MPT results with
      $B_\sigma = B_\sigma^{\rm (HF)} = 0$.  Thin (solid and dashed) lines:
      $T_1$ (zeros of $\chi^{-1}$).  Points with error bars: QMC results
      (zeros of $\chi^{-1}$) by Ulmke~\protect\cite{Ulm97}.}
    \end{center}
    \label{fig:fcc_Tc}
\end{figure}

In conclusion, ensuring the correctness of the moments of the spectral density
up to $n=3$, is a necessary condition to be consistent with the rigorous
results of Harris and Lange for the strong-coupling regime. In this respect
the modified perturbation theory not only represents a conceptual improvement
upon the approach of Kajueter and Kotliar, but also yields closer agreement
with QMC data. The (possible) spin-dependence of the higher-order correlation
functions ($B_\sigma$) that appear in the $n=3$ moment is important for
ferromagnetic order and has been shown to affect critical fillings and
temperatures considerably. On the contrary, there are only minor effects of
$B_\sigma$ for the paramagnet.  Improvement upon the IPT can also be expected
for the antiferromagnet at half-filling.
\\

This work has been supported by the Deutsche Forschungsgemeinschaft (SFB 290).
%
\begin{figure}
    \begin{center}
    \mbox{} \epsfxsize=0.9\linewidth \epsffile{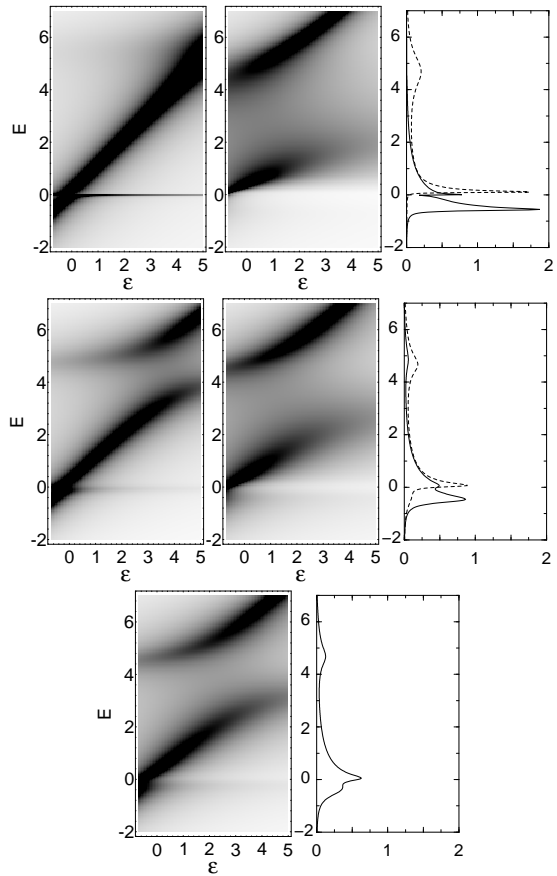}
    \caption{Density plots
      of the spectral density: energy on the y-axis, Bloch dispersion
      $\epsilon ({\bold k})$ on the x-axis. $U=4$, $n=0.6$ and the
      fcc-lattice. First row: $\uparrow$-spectral density,
      $\downarrow$-spectral density and density of states $\rho_\sigma(E-\mu)$
      (solid line $\uparrow$, dashed line: $\downarrow$) at $k_BT=0.022$
      (magnetisation $m=0.57$). Second row: $k_BT=0.058$, $m=0.32$. Third row:
      $k_BT=k_BT_C=0.065$.}
    \end{center}
    \label{fig:sd}
\end{figure}



\begin{references}

\bibitem{Hub63}
J. Hubbard, Proc. R. Soc. London, Ser. A {\bf 276},  283  (1963).

\bibitem{MV89}
W. Metzner and D. Vollhardt, Phys. Rev. Lett. {\bf 62},  324  (1989).

\bibitem{Vol93}
D. Vollhardt,  in {\em Correlated Electron Systems}, edited by V.~J. Emery
  (World Scientific, Singapore, 1993), p.\ 57.

\bibitem{GK92}
A. Georges and G. Kotliar, Phys. Rev. B {\bf 45},  6479  (1992).

\bibitem{Jar92}
M. Jarrell, Phys. Rev. Lett. {\bf 69},  168  (1992).

\bibitem{GK93}
A. Georges and W. Krauth, Phys. Rev. B {\bf 48},  7167  (1993).

\bibitem{KK96}
H. Kajueter and G. Kotliar, Phys. Rev. Lett. {\bf 77},  131  (1996).

\bibitem{HL67}
A.~B. Harris and R.~V. Lange, Phys. Rev. {\bf 157},  295  (1967).

\bibitem{NB89}
W. Nolting and W. Borgie{\l}, Phys. Rev. B {\bf 39},  6962  (1989).

\bibitem{PWN97}
M. Potthoff, T. Wegner, and W. Nolting, Phys. Rev. B {\bf 55},  16132  (1997).

\bibitem{JP93}
M. Jarrell and T. Pruschke, Z. Phys. B {\bf 90},  187  (1993).

\bibitem{PJF95}
T. Pruschke, M. Jarrell, and J.~K. Freericks, Adv. Phys. {\bf 44},  187
  (1995).

\bibitem{MHa89}
E. M{\"u}ller-Hartmann, Z. Phys. B {\bf 74},  507  (1989).

\bibitem{OPK97}
T. Obermeier, T. Pruschke, and J. Keller,  Phys. Rev. B {\bf 56}, 8479 (1997).

\bibitem{Ulm97}
M. Ulmke, preprint cond-mat 9704229  (1997).

\bibitem{VBHKSU97}
D. Vollhardt, N. Bl{\"u}mer, K. Held, M. Kollar, J. Schlipf, and M. Ulmke, Z.
  Phys. B {\bf 103},  283  (1997).

\bibitem{HC94}
E. Halvorsen and G. Czycholl, J. Phys.: Condens. Matter {\bf 6}, 10331 (1994)

\bibitem{NBB91}
W. Nolting, S. {Bei~der~Kellen}, and G. Borstel, Phys. Rev. B {\bf 43},  1117
  (1991).

\end{references}
\end{document}